\documentclass[conference]{IEEEtran}
\usepackage{amsmath}
\usepackage{amssymb}
\usepackage[table]{xcolor}
\usepackage{booktabs}
\usepackage{adjustbox}
\usepackage{amsfonts}
\usepackage{booktabs}
\usepackage{subcaption}
\usepackage{graphicx}
\usepackage[table]{xcolor}   
\usepackage{colortbl}
\usepackage{booktabs}   
\IEEEoverridecommandlockouts

\ifCLASSINFOpdf
\else
\fi
\begin{document}
%

\title{Efficient Temporal Modeling for Mobile Sleep Staging via Lightweight Random Attention}







\author{
\IEEEauthorblockN{Guisong Liu, Pengfei Wei\IEEEauthorrefmark{1}}
\IEEEauthorblockA{\textit{School of Biological Science and Medical Engineering} \\
\textit{Southeast University}\\
Nanjing, China \\
230258331@seu.edu.cn, 101014012@seu.edu.cn}

\IEEEauthorblockN{Jiansong Zhang\IEEEauthorrefmark{1}}
\IEEEauthorblockA{\textit{School of Computer Science \& Software Engineering} \\
\textit{Shenzhen University}\\
Shenzhen, China \\
2453103003@mails.szu.edu.cn}

\IEEEauthorblockN{Martin Dresler\IEEEauthorrefmark{1}}
\IEEEauthorblockA{\textit{Donders Institute for Brain, Cognition and Behaviour} \\
\textit{Radboud University Medical Center}\\
Nijmegen, The Netherlands\\
martin.dresler@donders.ru.nl}
\thanks{*Corresponding author.}
}



%


\maketitle
\begin{abstract}
Mobile sleep staging serves as a foundational infrastructure for in-home sleep monitoring and closed-loop modulation. But existing sequential models such as RNNs and Transformers are computationally expensive for mobile deployment. In this paper, we propose Random Attention (RA), a lightweight temporal modeling module based on fixed random projections, which replaces learnable sequence modeling with similarity-based aggregation. RA introduces little additional parameters beyond the epoch encoder while enabling effective temporal smoothing. We further provide a theoretical interpretation via the Random Attention Prior Kernel (RAPK), which decomposes RA into a global smoothing term and a feature similarity term, offering an interpretable view of temporal sleep structure. Experiments on Sleep-EDF-20 and Sleep-EDF-78 show that RA consistently improves epoch-wise baselines by 1–3\% in accuracy and F1 score, while achieving competitive performance compared with LSTM, GRU, and Transformer models. RA also demonstrates strong generalization across different backbone encoders and improved robustness over conventional temporal smoothing methods. These results indicate that efficient sleep staging can be achieved through lightweight similarity-based temporal aggregation, making RA suitable for real-time wearable applications.
\end{abstract}


%
\IEEEpeerreviewmaketitle

\section{Introduction}

Automatic sleep staging is fundamental to large-scale sleep health monitoring, digital phenotyping, and closed-loop neuromodulation~\cite{liu2025sleepmodulationchallengetransitioning,esfahani2023closed,10.3389/fnins.2025.1682450}. Although polysomnography (PSG) remains the clinical gold standard, its reliance on expensive equipment and labor-intensive expert annotation limits scalability. Recent advances in wearable sensing, particularly portable EEG devices, have enabled mobile sleep staging in home environments~\cite{coon_ezscore-f_2025,10.1093/sleepadvances/zpaf094,sikder2025eegflosspythonpackagerefining,Esfahani2023.08.18.553744}. These systems support long-term, real-world sleep monitoring and real-time closed-loop intervention. However, their deployment on mobile devices remains constrained by computational and energy budgets, limiting high-accuracy sleep staging in practice~\cite{10.3389/fnins.2023.1218072}.

Existing deep learning approaches can be broadly categorized into epoch-wise modeling and sequence modeling. Epoch-wise methods process each 30-second EEG epoch independently~\cite{10.3389/fnins.2023.1218072,wang2026ulwsleepnetultralightweightnetworkmultimodal,GOERTTLER2026110141}. While computationally efficient, they ignore the strong temporal continuity of sleep architecture, often resulting in unstable predictions that violate physiological transition patterns. To address this limitation, sequence models such as LSTM and GRU have been widely adopted to capture temporal dependencies across neighboring epochs~\cite{7961240,9176741,8631195}. More recently, Transformer-based models have shown improved performance by modeling long-range dependencies via self-attention~\cite{9697331,lee_explainable_2025}. However, these gains come at the cost of increased computational complexity, memory usage, and inference latency, making them less suitable for resource-constrained mobile applications. This motivates a closer look at whether such modeling complexity is fundamentally required for sleep staging.

Empirical evidence further suggests that extending the temporal context in conventional sequence models often yields only marginal or inconsistent performance gains~\cite{10623416,10210638,10782964,10782771}; if long-range dependency modeling were the primary driver, performance would be expected to improve systematically with larger temporal windows. This indicates that the benefits of temporal modeling may stem more from enforcing local temporal consistency than from capturing complex long-range interactions. This observation is consistent with the physiological characteristics of sleep, where stage transitions are typically smooth, gradual, and highly redundant, with neighboring epochs sharing similar patterns. Therefore, a fundamental question arises: is the assumption that sleep staging requires modeling complex long-range dependencies truly justified, or can the temporal structure be more effectively captured by simpler smoothing mechanisms grounded in the physiological continuity of sleep stage transitions?

Our previous work with stochastic transformers suggests that sleep staging can be effectively interpreted as an adaptive smoothing process, where stochastic attention suppresses local noise while preserving meaningful transitions based on feature similarity~\cite{liu2026rethinkingrandomtransformersadaptive}. This perspective implies that performance gains primarily arise from enforcing temporal consistency rather than learning complex dependencies.

Motivated by this insight, we propose a lightweight Random Attention (RA) mechanism for mobile sleep staging. Instead of learning parameter-intensive temporal dependencies, RA performs content-aware temporal aggregation using fixed random projections, achieving efficient sequence modeling with minimal computational overhead. This design explicitly leverages the physiological characteristics of sleep stage transitions, making it well-suited for real-time deployment on resource-constrained devices.

Extensive experiments on benchmark datasets demonstrate that the proposed method achieves performance comparable to conventional sequence models while significantly reducing model size and computational cost. It also outperforms standard post-processing smoothing methods in both robustness and peak performance.

The main contributions of this paper are summarized as follows:
\begin{itemize}
    \item We revisit temporal sleep staging from the perspective of adaptive smoothing, challenging the necessity of complex dependency modeling.
    \item We propose a lightweight Random Attention mechanism that enables efficient, content-aware temporal modeling.
    \item We demonstrate through extensive experiments that the proposed method achieves competitive performance with improved efficiency and robustness compared to both sequence models and conventional smoothing approaches.
\end{itemize}

\section{Method}

\subsection{Problem Definition}
Given a sequence of EEG epochs $X=\{x_1,x_2,\dots,x_T\}$, where each $x_t \in \mathbb{R}^{C \times L}$ is a 30-second segment ($L=3000$ at 100 Hz), the goal is to predict the sequence of sleep stage labels $Y=\{y_1,y_2,\dots,y_T\}$, $y_t \in \{W, N1, N2, N3, REM\}$.

Most mobile systems first extract epoch-level representations $Z=\{z_1,\dots,z_T\}$, $z_t \in \mathbb{R}^d$, via a lightweight CNN encoder, then apply temporal modeling. We replace costly LSTM/GRU/Transformer modules with a Random Attention (RA).

\subsection{Random Attention}

RA constructs a lightweight random attention matrix $A$ and aggregates features as $O = AZ$, where $A$ is never learned. Instead, each epoch is projected into a fixed random low-dimensional space:
\[
Q = Z W_Q, \quad K = Z W_K, \quad W_Q, W_K \in \mathbb{R}^{d \times d_k}
\]
with entries sampled once at initialization and kept frozen thereafter. The attention weights are then~\cite{vaswani2023attentionneed}
\[
A = \text{softmax}\left(\frac{QK^\top}{\sqrt{d_k}}\right)
\]
(row-wise). Since $W_Q$ and $W_K$ are fixed, RA introduces almost no additional trainable parameters, and only the epoch encoder and final classifier are learned.

Based on our prior study, the projection matrices are initialized using Xavier uniform initialization~\cite{pmlr-v9-glorot10a}.
\[
W \sim \mathcal{U}\left(-\sqrt{\frac{6}{d + d_k}}, \, \sqrt{\frac{6}{d + d_k}}\right).
\]
 Empirically, uniform schemes (e.g., Xavier or Kaiming) outperform Gaussian alternatives. From a kernel perspective, Gaussian initialization compresses feature variance, causing attention logits to collapse toward zero and degenerating the model into near-uniform averaging. In contrast, uniform initialization preserves feature scale, keeping attention scores in a stable regime that balances smoothing and structure preservation, which is critical for capturing temporal dependencies in sleep stage transitions.

\subsection{Theoretical Interpretation}
The effectiveness of RA follows directly from the Random Attention Prior Kernel (RAPK) established in our prior work~\cite{liu2026rethinkingrandomtransformersadaptive}. In the high-dimensional limit ($d_k \rightarrow \infty$), the expected kernel converges to
\[
\mathbb{E}[K_{\text{RAP}}] \approx C_0 \mathbf{1}\mathbf{1}^\top + C_1 ZZ^\top,
\]
where $ZZ^\top$ is the Gram matrix of epoch representations, and $C_0, C_1$ are positive, sequence-dependent scalars, whose magnitudes are determined by the initialization variances and sequence length, and scale proportionally with the global statistics of the input features.

This decomposition reveals that RA implements content-aware temporal smoothing:

\begin{itemize}
    \item The global term $C_0 \mathbf{1}\mathbf{1}^\top$ enforces temporal inertia and suppresses high-frequency noise.
    \item The similarity term $C_1 ZZ^\top$ adaptively weights interactions according to feature similarity.
\end{itemize}

Consequently, the RAPK kernel combines a global averaging term and a content-adaptive smoothing term, matching the physiology of sleep staging.

The global term $C_0 \mathbf{1}\mathbf{1}^\top$ captures sleep-state inertia: sleep changes slowly, so adjacent epochs are likely to share the same stage. It therefore applies a uniform averaging bias within the local window, suppressing isolated fluctuations and noisy predictions.

However, uniform averaging alone would blur true stage transitions. The content-adaptive term $C_1 ZZ^\top$ adjusts the smoothing strength according to feature similarity: epochs with similar representations are smoothed together, whereas epochs with dissimilar representations, typically those located across stage boundaries, interact much more weakly. As a result, RAPK enforces strong smoothing within a stage while preserving meaningful transitions between stages.

\subsection{Computational Complexity}

Standard self-attention has quadratic complexity in sequence length, requiring $\mathcal{O}(T^2 d + T d^2)$ computation and $\mathcal{O}(T^2 + T d)$ memory, which becomes prohibitive for long EEG sequences. 

In contrast, RA replaces the explicit pairwise attention computation with a fixed random projection mechanism, eliminating the need to construct the full $T \times T$ attention matrix. This reduces the computational cost to $\mathcal{O}(T d D_k)$ and memory cost to $\mathcal{O}(T D_k)$. The absence of the feed-forward network further reduces the overall complexity while preserving temporal interaction through low-rank token mixing.

The resulting structured projection avoids dense attention storage and enables full parallelization across time. Compared with LSTM, GRU, and Transformer baselines, RA achieves substantially lower latency and memory consumption, making it well-suited for real-time wearable EEG inference scenarios. Table~\ref{tab:complexity} quantitatively compares the computational and memory complexity of different models.

\begin{table}[h]
\centering
\caption{Computational complexity comparison. $T$ is sequence length, $d$ is feature dimension, and $D_k$ is the random projection dimension in RA.}
\begin{tabular}{lcc}
\hline
Model & Time Complexity (per sequence) & Memory Complexity \\
\hline
LSTM/GRU & $\mathcal{O}(T d^2)$ & $\mathcal{O}(T d)$ \\
Transformer & $\mathcal{O}(T^2 d + T d^2)$ & $\mathcal{O}(T^2 + T d)$ \\
RA (ours) & $\mathcal{O}(T d D_k)$ & $\mathcal{O}(T D_k)$ \\
\hline
\end{tabular}
\label{tab:complexity}
\end{table}

\begin{table*}[h]
\centering
\caption{Statistics of Sleep-EDF datasets.}
\label{tab:sleepedf}
\begin{tabular}{lcccccccccc}
\hline
Dataset & Subjects & Channel & Sampling Rate & Total & W & N1 & N2 & N3 & REM \\
\hline
Sleep-EDF-20 & 20 & Fpz-Cz & 100 Hz & 42308
& \shortstack{8285 \\ (19.6\%)}
& \shortstack{2804 \\ (6.6\%)}
& \shortstack{17799 \\ (42.1\%)}
& \shortstack{5703 \\ (13.5\%)}
& \shortstack{7717 \\ (18.2\%)} \\
\hline
Sleep-EDF-78 & 78 & Fpz-Cz & 100 Hz & 195479
& \shortstack{65951 \\ (33.7\%)}
& \shortstack{21522 \\ (11.0\%)}
& \shortstack{69132 \\ (35.4\%)}
& \shortstack{13039 \\ (6.7\%)}
& \shortstack{25835 \\ (13.2\%)} \\
\hline
\end{tabular}
\label{tab:sleepedf}
\end{table*}

\section{EXPERIMENTS}

\subsection{Datasets}

We evaluate on Sleep-EDF-20 and Sleep EDFX~\cite{doi:10.1161/01.CIR.101.23.e215}\cite{867928}. Both contain overnight EEG recordings (Fpz-Cz channel, 100 Hz) annotated into five stages (W, N1, N2, N3, REM). We follow standard subject-independent cross-validation after excluding movement/unknown epochs and merging S3/S4 into N3 per American Academy of Sleep Medicine guidelines. Following prior work, only 30 minutes of wakefulness before sleep onset and after sleep termination are retained~\cite{9417097}. For Sleep-EDF-20, we adopt 20-fold cross-validation, while for Sleep-EDFX, we adopt 10-fold cross-validation for consistency with previous studies. Table~\ref{tab:sleepedf} summarizes the statistics of the datasets used in this study.

\subsection{Implementation Details}

We adopt a lightweight CNN-based epoch encoder followed by a temporal modeling module for sleep staging. Specifically, we build upon our previous mobile sleep staging model, MicrosleepNet~\cite{10.3389/fnins.2023.1218072}, which consists of two components: (1) a group-convolution-based feature extraction encoder and (2) a dilated-convolution-based feature fusion module. We consider two backbone settings. The first, denoted as MicrosleepNet\_Encoder, contains only the feature extraction encoder. The second, denoted as MicrosleepNet, includes the complete original architecture. For RA, the default random projection dimension is $d_k = 128$. Baselines include: (i) epoch encoder only, (ii) LSTM~\cite{10.1162/neco.1997.9.8.1735}, (iii) GRU~\cite{chung2014empiricalevaluationgatedrecurrent}, and (iv) a trainable Transformer~\cite{vaswani2023attentionneed}. Each sample consists of a sliding window of 10 consecutive epochs.

Training is performed for 100 epochs using AdamW with an initial learning rate of $1\times10^{-3}$, weight decay of $1\times10^{-4}$, and batch size 20. For the trainable Transformer baseline, Transformer layers and learnable positional embeddings are optimized with a reduced learning rate of $1\times10^{-4}$. We apply early stopping with a patience of 10 epochs, a 5-epoch warmup schedule, and gradient clipping with a maximum norm of 2.0. Evaluation is conducted using overall accuracy, weighted F1, Cohen's kappa, and per-stage F1.

All experiments are implemented in PyTorch and conducted on a single NVIDIA RTX 3090 GPU (24GB). No additional signal preprocessing, data augmentation, or class balancing strategies are applied to ensure a fair comparison across models.

\begin{table*}[t]
\small
\setlength{\tabcolsep}{3.5pt}
\renewcommand{\arraystretch}{1.10}
\begin{adjustbox}{width=\textwidth}
\begin{tabular}{lccccccccccc}
\toprule
Model & Acc (\%) & F1 (\%) & Kappa & W & N1 & N2 & N3 & REM & Params & MFLOPs \\
\midrule

\multicolumn{11}{c}{\textit{Sleep-EDF-20}} \\
\midrule

\rowcolor{gray!20}
MicroSleepNet\_Encoder & 81.79 & 81.62 & 0.7496 & 0.8934 & 0.3238 & 0.8648 & 0.8799 & 0.7546 & \textbf{4.77K} & \textbf{21.17} \\
\rowcolor{white}
\textbf{MicroSleepNet\_Encoder\_RA} & \textbf{83.35 (+1.56)} & \textbf{83.23 (+1.61)} & \textbf{0.7712 (+0.02)} & \textbf{0.8670 (-0.03)} & \textbf{0.4281 (+0.10)} & \textbf{0.8750 (+0.01)} & \textbf{0.8828 (+0.00)} & \textbf{0.8070 (+0.05)} & \underline{71.07K} & \underline{21.86} \\
MicroSleepNet\_Encoder\_trainT & 83.85 & 83.71 & 0.7781 & 0.8731 & 0.4432 & 0.8734 & 0.8775 & 0.8287 & 204.32K & 23.16 \\
MicroSleepNet\_Encoder\_LSTM & 83.82 & 83.82 & 0.7782 & 0.8989 & 0.4301 & 0.8713 & 0.8779 & 0.8167 & 136.87K & 22.48 \\
MicroSleepNet\_Encoder\_GRU & 83.90 & 83.99 & 0.7786 & 0.8896 & 0.4478 & 0.8782 & 0.8831 & 0.8097 & 103.84K & 22.15 \\

\rowcolor{gray!20}
MicroSleepNet & 82.36 & 82.10 & 0.7569 & 0.8902 & 0.3320 & 0.8716 & 0.8821 & 0.7642 & \textbf{48.23K} & \textbf{37.25} \\
\rowcolor{white}
\textbf{MicroSleepNet\_RA} & \textbf{83.70 (+1.34)} & \textbf{83.78 (+1.68)} & \textbf{0.7772 (+0.02)} & \textbf{0.8858 (0.00)} & \textbf{0.4773 (+0.15)} & \textbf{0.8708 (0.00)} & \textbf{0.8683 (-0.01)} & \textbf{0.8192 (+0.06)} & \underline{114.53K} & \underline{37.94} \\
MicroSleepNet\_trainT & 84.34 & 84.16 & 0.7850 & 0.8907 & 0.4674 & 0.8725 & 0.8785 & 0.8271 & 247.78K & 39.24 \\
MicroSleepNet\_LSTM & 84.17 & 84.01 & 0.7823 & 0.8847 & 0.4590 & 0.8765 & 0.8815 & 0.8170 & 180.32K & 38.56 \\
MicroSleepNet\_GRU & 83.89 & 83.84 & 0.7793 & 0.8959 & 0.4555 & 0.8760 & 0.8706 & 0.8060 & 147.29K & 38.23 \\

\midrule
\multicolumn{11}{c}{\textit{Sleep-EDF-78}} \\
\midrule

\rowcolor{gray!20}
MicroSleepNet\_Encoder & 78.32 & 76.99 & 0.6972 & 0.9044 & 0.3033 & 0.8240 & 0.8025 & 0.6576 & \textbf{4.77K} & \textbf{21.17} \\
\rowcolor{white}
\textbf{MicroSleepNet\_Encoder\_RA} & \textbf{80.75 (+2.43)} & \textbf{79.93 (+2.94)} & \textbf{0.7334 (+0.04)} & \textbf{0.9199 (+0.02)} & \textbf{0.3715 (+0.07)} & \textbf{0.8355 (+0.01)} & \textbf{0.7974 (-0.01)} & \textbf{0.7548 (+0.10)} & \underline{71.07K} & \underline{21.86} \\
MicroSleepNet\_Encoder\_trainT & 81.20 & 80.68 & 0.7396 & 0.9162 & 0.4252 & 0.8380 & 0.7908 & 0.7728 & 204.32K & 23.16 \\
MicroSleepNet\_Encoder\_LSTM & 80.86 & 80.51 & 0.7352 & 0.9168 & 0.4202 & 0.8413 & 0.7982 & 0.7501 & 136.87K & 22.48 \\
MicroSleepNet\_Encoder\_GRU & 80.46 & 79.86 & 0.7289 & 0.9118 & 0.3972 & 0.8379 & 0.7916 & 0.7450 & 103.84K & 22.15 \\

\rowcolor{gray!20}
MicroSleepNet & 79.20 & 78.43 & 0.7113 & 0.9080 & 0.3488 & 0.8311 & 0.7987 & 0.7017 & \textbf{48.23K} & \textbf{37.25} \\
\rowcolor{white}
\textbf{MicroSleepNet\_RA} & \textbf{81.19 (+1.99)} & \textbf{80.78 (+2.35)} & \textbf{0.7398 (+0.03)} & \textbf{0.9145 (+0.01)} & \textbf{0.4109 (+0.06)} & \textbf{0.8369 (+0.01)} & \textbf{0.8033 (+0.00)} & \textbf{0.7929 (+0.09)} & \underline{114.53K} & \underline{37.94} \\
MicroSleepNet\_trainT & 81.83 & 81.28 & 0.7481 & 0.9186 & 0.4236 & 0.8420 & 0.8048 & 0.7952 & 247.78K & 39.24 \\
MicroSleepNet\_LSTM & 81.68 & 81.13 & 0.7460 & 0.9136 & 0.4383 & 0.8443 & 0.8010 & 0.7802 & 180.32K & 38.56 \\
MicroSleepNet\_GRU & 81.14 & 80.76 & 0.7392 & 0.9121 & 0.4363 & 0.8397 & 0.7970 & 0.7717 & 147.29K & 38.23 \\

\bottomrule
\end{tabular}
\end{adjustbox}
\caption{Main results. Baselines are highlighted in gray, and proposed RA variants are fully bolded. Improvements over baselines are shown in parentheses. Computational cost (Params and MFLOPs) is added.}
\label{tab:Main Results}
\end{table*}

\subsection{Main Results}
Table~\ref{tab:Main Results} summarizes the full performance comparison of the proposed Random Attention (RA) against the epoch-wise baseline and three strong sequence modeling baselines on both Sleep-EDF-20 and Sleep-EDFX. Results are reported for two MicroSleepNet variants: MicroSleepNet\_Encoder and MicroSleepNet model.

On Sleep-EDF-20, the plain MicroSleepNet\_Encoder baseline achieves 81.79\% accuracy and 81.62\% weighted F1. Adding the RA module improves performance to 83.35\% accuracy (+1.56\%) and 83.23\% weighted F1 (+1.61\%). A similar trend is observed on the full MicroSleepNet model, where RA increases accuracy from 82.36\% to 83.70\% and weighted F1 from 82.10\% to 83.78\%. These improvements are achieved with negligible additional trainable parameters. Compared with other temporal modeling variants, RA achieves comparable performance, with results generally lying within the same performance range as LSTM and GRU-based designs, while remaining slightly below the best-performing LSTM configuration. It also remains below the best-performing trainable Transformer, which achieves 84.34\% accuracy and 84.16\% weighted F1. Overall, RA provides a favorable trade-off between performance and efficiency, delivering consistent gains over encoder-only baselines with minimal computational overhead.

The trend is even more pronounced on the larger and more challenging Sleep-EDFX dataset. The MicroSleepNet\_Encoder baseline yields only 78.32\% accuracy and 76.99\% weight F1. RA improves these by +2.43\% and +2.94\%, respectively, reaching 80.75\% accuracy and 79.93\% weight F1. For the full MicroSleepNet model, RA lifts accuracy from 79.20\% to 81.19\% and weight F1 from 78.43\% to 80.78\%. On this dataset, RA achieves performance comparable to LSTM and GRU, and approaches the trainable Transformer (81.83\% accuracy).These consistent improvements across both backbones and both datasets demonstrate that the lightweight random attention mechanism delivers highly effective temporal smoothing that rivals or exceeds far more expensive sequence models.

\subsection{Analysis of Different Sleep Stages}
The per-stage F1 scores in Table~\ref{tab:Main Results} reveal that RA’s benefits are selective and clinically meaningful. The largest and most consistent improvements occur on the challenging transitional stages N1 and REM, while performance on the more stable stages (W, N2, N3) remains strong or shows modest gains. This pattern holds across both datasets and both MicroSleepNet backbones.

On both datasets, RA yields its largest improvements on N1, increasing F1 by approximately 0.06--0.15 depending on the backbone. REM also benefits substantially, with gains ranging from about 0.04 to 0.10. In contrast, the more stable stages (W, N2, and N3) exhibit only modest improvements, typically around 0.01. These results align directly with the Random Attention Prior Kernel (RAPK) analysis: N1 epochs are short, unstable, and frequently confused with neighboring stages, yet they remain close in feature space to correct neighbors. The similarity term $C_1 ZZ^\top$ selectively strengthens these interactions, correcting noisy predictions without explicit transition modeling. REM benefits analogously due to shared transitional EEG patterns. Meanwhile, the global averaging term $C_0 \mathbf{1}\mathbf{1}^\top$ provides sufficient regularization for stable stages without over-smoothing discriminative boundaries. Overall, RA reinforces the natural temporal continuity of sleep physiology, delivering competitive macro-level performance at a fraction of the computational cost of conventional sequence models.

\subsection{Effect of Random Projection Dimension $d_k$}
We further ablate the random projection dimension $d_k$ (default 128) on Sleep-EDF-20. Results are shown in Table~\ref{tab:Ablation}. Performance improves monotonically with larger $d_k$, reaching a peak of 84.28\% Acc at $d_k=512$. This trend is theoretically predicted by RAPK: higher dimensionality yields a more accurate low-variance approximation of the similarity kernel $C_1 ZZ^\top$, resulting in more stable content-aware smoothing. However, excessively strong smoothing can slightly reduce sensitivity to transient stages such as N1. This indicates a trade-off between prediction smoothness and sensitivity to short transitional stages, which can be controlled by adjusting $d_k$. Even at the modest default $d_k=128$, RA already delivers strong gains, making it highly practical for mobile deployment.

\begin{table*}[t]
\centering
\caption{Ablation on random projection dimension $d_k$ (MicroSleepNet backbone, Sleep-EDF-20).}
\begin{tabular}{lcccccccc}
\hline
Model & Acc (\%) & F1 (\%) & Kappa & W & N1 & N2 & N3 & REM \\
\hline
MicroSleepNet\_Encoder\_RA (128) & \underline{83.35} & 83.23 & 0.7712 & 0.8670 & \underline{0.4281} & \textbf{0.8750} & \textbf{0.8828} & 0.8070 \\
MicroSleepNet\_Encoder\_RA (256) & 83.34 & \underline{83.38} & \underline{0.7715} & \underline{0.8832} & 0.4098 & 0.8699 & 0.8759 & \underline{0.8241} \\
MicroSleepNet\_Encoder\_RA (512) & \textbf{84.28} & \textbf{84.16} & \textbf{0.7839} & \textbf{0.8975} & 0.4079 & \underline{0.8741} & \underline{0.8773} & \textbf{0.8390} \\

MicroSleepNet\_RA (128) & 83.70 & \underline{83.78} & 0.7772 & \textbf{0.8858} & \textbf{0.4773} & 0.8708 & 0.8683 & 0.8192 \\
MicroSleepNet\_RA (256) & \underline{84.19} & 83.83 & \underline{0.7821} & 0.8716 & 0.4356 & \textbf{0.8767} & \underline{0.8750} & \textbf{0.8341} \\
MicroSleepNet\_RA (512) & \textbf{84.28} & \textbf{84.02} & \textbf{0.7838} & \underline{0.8785} & \underline{0.4571} & \underline{0.8757} & \textbf{0.8842} & \underline{0.8244} \\
\hline
\end{tabular}
\label{tab:Ablation}
\end{table*}

\definecolor{groupblue}{RGB}{235,242,250}
\definecolor{groupgreen}{RGB}{236,247,236}
\definecolor{grouporange}{RGB}{252,244,232}
\definecolor{grouppurple}{RGB}{243,238,250}

\begin{table*}[t]
\centering
\caption{Generalization across epoch encoders (Sleep-EDF-20).}
\begin{tabular}{lcccccccc}
\hline
Model & Acc (\%) & F1 (\%) & Kappa & W & N1 & N2 & N3 & REM \\
\hline

\rowcolor{groupblue}
DeepSleepNet\_Encoder & 82.07 & 81.40 & 0.7531 & \underline{0.8577} & \underline{0.2858} & \underline{0.8744} & \textbf{0.8831} & \underline{0.7697} \\

\rowcolor{groupblue}
DeepSleepNet & \underline{84.06} & \underline{83.75} & \underline{0.7812} & \underline{0.8604} & \underline{0.4319} & \underline{0.8793} & \underline{0.8661} & \underline{0.8431} \\

\rowcolor{groupblue}
DeepSleepNet\_Encoder\_RA & \textbf{84.92} & \textbf{84.63} & \textbf{0.7931} & \textbf{0.8738} & \textbf{0.4405} & \textbf{0.8827} & \underline{0.8934} & \textbf{0.8462} \\

\rowcolor{groupgreen}
TinySleepNet\_Encoder & 81.60 & 80.92 & 0.7465 & \underline{0.8567} & \underline{0.2600} & \underline{0.8735} & \textbf{0.8828} & \underline{0.7560} \\

\rowcolor{groupgreen}
TinySleepNet & \underline{83.17} & \underline{82.80} & \underline{0.7679} & \underline{0.8545} & \underline{0.3986} & \underline{0.8763} & \underline{0.8687} & \underline{0.8148} \\

\rowcolor{groupgreen}
TinySleepNet\_Encoder\_RA & \textbf{83.75} & \textbf{83.46} & \textbf{0.7766} & \textbf{0.8871} & \textbf{0.4111} & \textbf{0.8763} & \underline{0.8423} & \textbf{0.8315} \\

\rowcolor{grouporange}
ULW\_SleepNet & \underline{80.76} & \underline{80.08} & \underline{0.7347} & \underline{0.8570} & \underline{0.2556} & \underline{0.8690} & \underline{0.8680} & \underline{0.7324} \\

\rowcolor{grouporange}
ULW\_SleepNet\_RA & \textbf{83.11} & \textbf{82.67} & \textbf{0.7680} & \textbf{0.8702} & \textbf{0.3859} & \textbf{0.8714} & \textbf{0.8739} & \textbf{0.8030} \\

\rowcolor{grouppurple}
MSA\_CNN (Large) & \underline{82.10} & \underline{81.63} & \underline{0.7536} & \underline{0.8610} & \underline{0.3266} & \underline{0.8743} & \textbf{0.8858} & \underline{0.7619} \\

\rowcolor{grouppurple}
MSA\_CNN\_RT (Large) & \textbf{84.20} & \textbf{84.04} & \textbf{0.7828} & \textbf{0.8867} & \textbf{0.4259} & \textbf{0.8818} & \underline{0.8828} & \textbf{0.8152} \\

\hline
\end{tabular}
\label{tab:Generalization}
\end{table*}

\begin{figure*}[!t]
\centering
\includegraphics[width=0.48\linewidth]{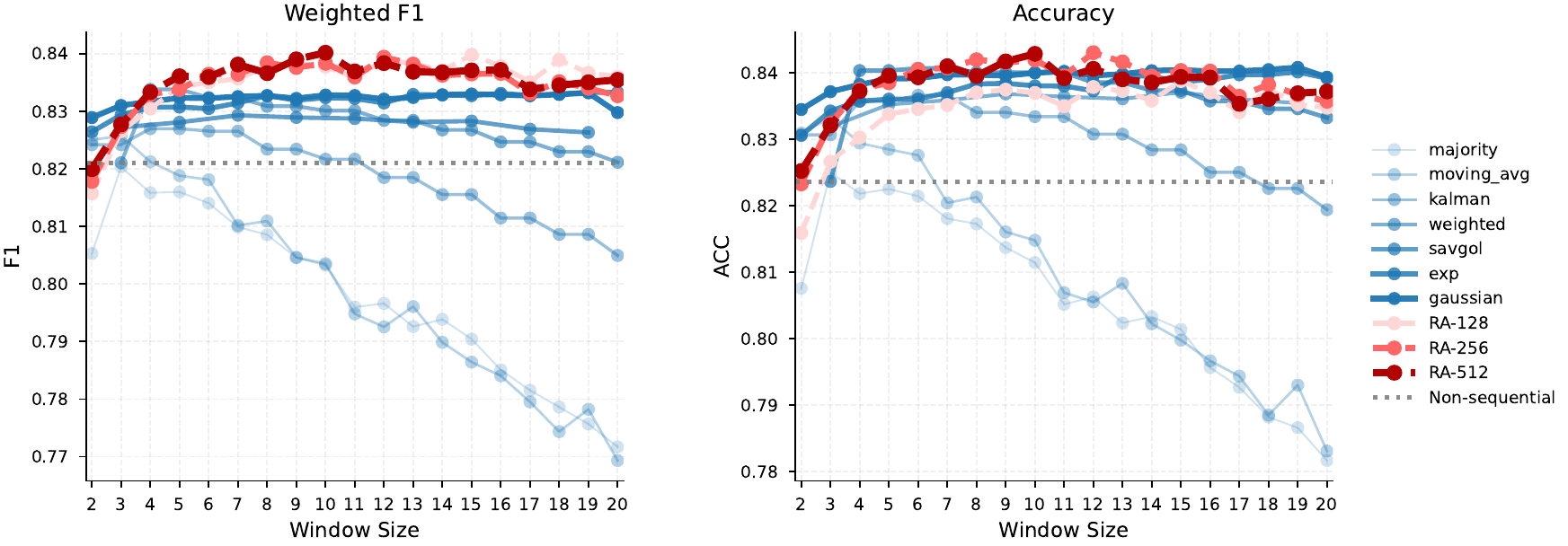}
\hfill
\includegraphics[width=0.48\linewidth]{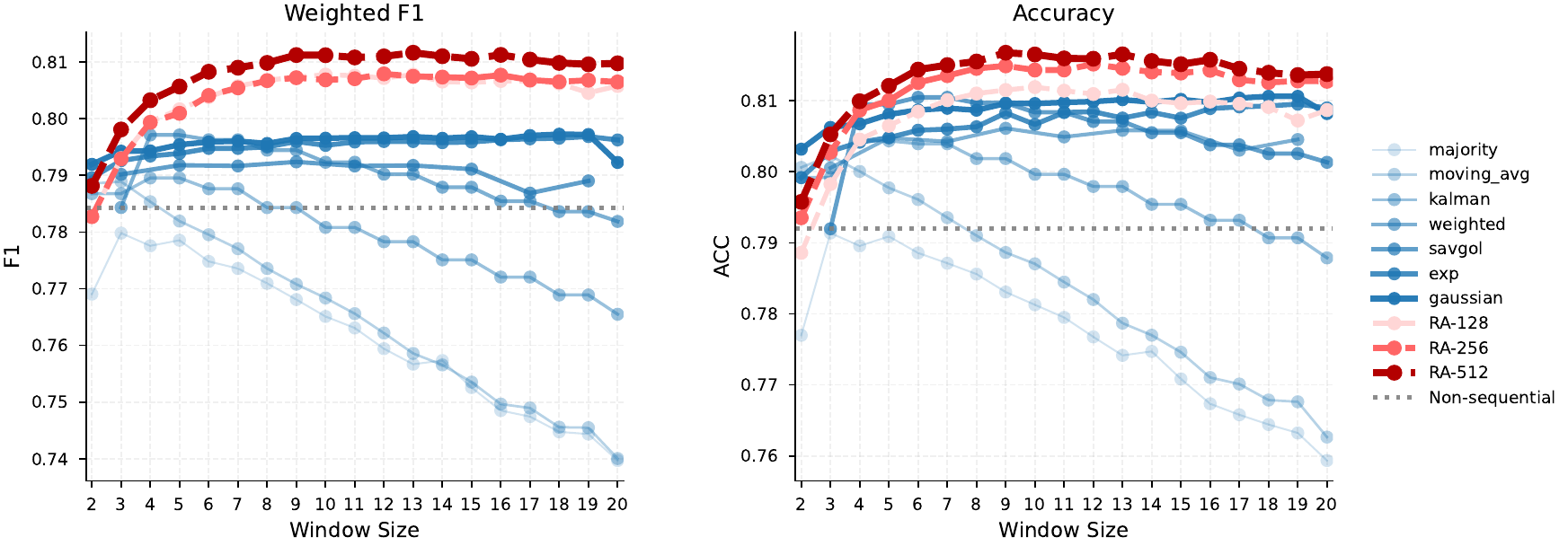}
\caption{Comparison with different window sizes on Sleep-EDF-20(left) and Sleep-EDF-78(right).}
\label{fig:window_size}
\end{figure*}

\subsection{Generalization Across Different Epoch Encoders}

To verify that RA is not tailored to a specific backbone, we evaluate it with four additional lightweight sleep staging encoders: DeepSleepNet~\cite{7961240}, TinySleepNet~\cite{9176741}, ULW-SleepNet~\cite{wang2026ulwsleepnetultralightweightnetworkmultimodal}, and MSA-CNN~\cite{GOERTTLER2026110141}. For DeepSleepNet, we adopted the implementation from LPSGM and further enhanced feature extraction by expanding the output channels of the final convolutional layer to 256~\cite{deng2024lpsgm}.

As shown in Table~\ref{tab:Generalization}, RA delivers consistent 2--3\% absolute improvements in accuracy and weighted F1 across all backbones, with especially large gains in N1 F1 (e.g., DeepSleepNet\_Encoder: +0.1547 and TinySleepNet\_Encoder: +0.1511). \textbf{Notably, the lightweight RA-enhanced variants even outperform the original DeepSleepNet and TinySleepNet architectures equipped with their native sequence modeling modules.} For example, DeepSleepNet\_Encoder\_RA surpasses the original DeepSleepNet by +0.86\% accuracy and +0.88\% weighted F1, while TinySleepNet\_Encoder\_RA also exceeds TinySleepNet despite using a substantially lighter temporal modeling design. These findings suggest that RA can capture sleep transition dynamics more effectively than conventional recurrent sequence modeling, while maintaining significantly lower computational complexity. Overall, the results establish RA as a plug-and-play, encoder-agnostic temporal smoothing module that can be seamlessly integrated into existing mobile sleep staging pipelines.

\subsection{Comparison with Simple Temporal Smoothing and Different Window Sizes}
We further compare RA with several non-learnable temporal smoothing methods on Sleep-EDF-20 and Sleep-EDF-78, including majority voting, moving average, Kalman filtering, weighted averaging, Savitzky--Golay filtering, and Gaussian smoothing. All methods are applied to the same epoch-level predictions while varying the temporal window from 2 to 20 epochs. For smoothing methods involving hyperparameters, we first perform grid search under a fixed window setting to identify the optimal configuration, ensuring that each baseline is evaluated under its best achievable performance.

As shown in Figure~\ref{fig:window_size}, introducing even simple temporal smoothing consistently improves all methods over the non-sequential baseline, confirming that local temporal context is beneficial for sleep staging. Across both datasets, most conventional smoothing methods achieve their best performance at moderate window sizes (typically 5--10 epochs), after which performance saturates or declines. The degradation is more pronounced for majority voting and Kalman filtering, where both F1 and accuracy decrease steadily with larger windows, while Gaussian and Savitzky--Golay filters remain relatively more robust but still exhibit mild performance drops beyond approximately 10 epochs.
On Sleep-EDF-78, a consistent but slightly lower overall performance level is observed compared with EDF-20, while the relative trends across methods remain highly consistent. In particular, the sensitivity of traditional smoothing approaches to increasing window size is still evident, with similar saturation behavior and subsequent degradation under overly large contexts. In contrast, RA maintains stable performance across a wide range of window sizes on both datasets. RA-128 already achieves competitive results, and increasing the projection dimension to 256 or 512 yields consistent improvements, with RA-512 achieving the best overall performance. Compared with strong conventional smoothing methods, RA not only attains higher peak performance on both EDF-20 and EDF-78, but also demonstrates markedly reduced sensitivity to window size selection. This behavior aligns with the RAPK interpretation, where conventional methods rely on fixed temporal weighting schemes, whereas RA adaptively aggregates information based on representation similarity.

\section{Discussion}

RA consistently improves both Sleep-EDF-20 and Sleep-EDF-78 by 1--3\% over the epoch-wise baseline while introducing almost no additional trainable parameters. Despite its simplicity, RA shows competitive performance compared with a fully trainable Transformer, while maintaining a modest performance gap. It also often matches or outperforms LSTM and GRU. This suggests that, for sleep staging, a substantial portion of useful temporal information can be effectively captured through lightweight similarity-based aggregation, without relying on complex sequence modeling. This behavior agrees with the RAPK interpretation. RA combines a global smoothing term, which suppresses isolated prediction noise, with a similarity term, which propagates information between epochs that have similar EEG representations. This simple content-aware smoothing is sufficient to capture the dominant temporal structure. The largest gains occur for the difficult transitional stages N1 and REM. These stages are often confused locally, but their neighboring epochs usually remain close in feature space. RA therefore improves them substantially by reinforcing consistent context. Finally, RA provides similar improvements across multiple backbones and random projection dimensions, demonstrating that it is a lightweight, plug-and-play temporal modeling module suitable for real-time wearable temporal sleep staging.

\section{Conclusion}

This paper demonstrates that the primary requirement for mobile sleep staging is lightweight, similarity-based temporal smoothing rather than complex sequence reasoning. We introduce RA, a random attention mechanism grounded in Random Transformer and RAPK theory. By constructing a structured smoothing kernel that balances global averaging and feature similarity, RA achieves efficient temporal aggregation. Experiments across multiple datasets and encoders confirm that RA achieves performance comparable to traditional temporal modeling baselines while delivering superior efficiency. Compared with a series of simple smoothing baselines, RA is also more stable and achieves higher peak performance. The practical significance of RA lies in its ability to provide accurate temporal sleep staging directly on wearable devices with negligible computational overhead, thereby making long-term home monitoring and future sleep closed-loop modulation systems feasible.

\section*{Acknowledgment}

The authors gratefully acknowledge the Brain-Computer Modulation Laboratory of Southeast University for providing the computational resources used in this study. The authors also gratefully acknowledge the financial support provided by the China Scholarship Council.



%




\bibliographystyle{IEEEtran}
\bibliography{ref}

\end{document}